\documentclass[12pt]{JHEP3}

\usepackage[utf8x]{inputenc}
\usepackage{amsfonts,amssymb,amsmath,amsthm,bbold,url,graphicx,psfrag}
\usepackage{mathrsfs}

\numberwithin{equation}{section}

\DeclareMathOperator{\SO}{SO}

\DeclareMathOperator{\SU}{SU}
\DeclareMathOperator{\U}{U}

\DeclareMathOperator{\shs}{shs}
\DeclareMathOperator{\hs}{hs}

\newcommand{\be}{ \begin{equation}}
\newcommand{\ee}{\end{equation}}

\newcommand{\sWinf}{\mathscr{W}_{\infty}}
\newcommand{\sW}{\mathscr{W}}

\title{Duality in ${\cal N}=2$ minimal model holography}
\author{Constantin Candu and Matthias R.\ Gaberdiel \\
Institut f\"ur Theoretische Physik , ETH Z\"urich \\
CH-8093 Z\"urich, Switzerland}

\abstract{Recently a duality between a  family of  ${\cal N}=2$ supersymmetric 
higher spin theories on AdS$_3$, and the  't~Hooft like limit of a class of Kazama-Suzuki 
models (that are parametrised by $N$ and $k$) was  proposed. The higher spin theories 
can be described by a
Chern-Simons theory based on the infinite-dimensional Lie algebra
${\rm shs}[\mu]$, and under the duality, $\mu$ is to be identified with
$\lambda=\frac{N}{N+k+1}$. Here we elucidate the 
structure of the (quantum) asymptotic symmetry algebra $\sW_{\infty}[\mu]$
for arbitrary $\mu$ and central charge $c$. In particular, we show that for each 
value of the central charge,
there are generically four different values of $\mu$ that describe the same $\sW_{\infty}$ 
algebra. Among other things this proves that the quantum symmetries on both
sides of the duality agree; this equivalence  does not just hold in the 't~Hooft limit, but 
even at finite $N$ and $k$.}

\begin{document}

%



\section{Introduction and Summary}

During the last few years dualities between higher spin theories on AdS$_{d+1}$
and  weakly coupled $d$-dimensional conformal field theories have
attracted  a lot of attention. The original idea that dualities of this kind should appear in 
the free field theory 
limit of the usual AdS/CFT correspondence was already noted some time ago
\cite{Sundborg:2000wp,Witten,Mikhailov:2002bp,Sezgin:2002rt}. However, 
a concrete proposal was only made by Klebanov \& Polyakov  \cite{Klebanov:2002ja}
(and generalised shortly afterwards by Sezgin \& Sundell  \cite{Sezgin:2003pt})
who suggested that the large $N$ limit of the ${\rm O}(N)$ vector model in $3$ dimensions
is dual to Vasiliev's (parity preserving) 
higher spin gravity on AdS$_4$ \cite{Vasiliev:1995dn,Vasiliev:1999ba}.
More recently, highly non-trivial evidence in favour of this proposal has been obtained
by comparing correlation functions of the two theories
\cite{Giombi:2009wh, Giombi:2010vg}. By now, the structure of these correlation
functions has been understood conceptually \cite{Maldacena:2011jn,Maldacena:2012sf},
and various further generalisations (in particular to parity violating theories) have been proposed 
and studied 
\cite{Aharony:2011jz,Giombi:2011kc,Chang:2012kt,Aharony:2012nh,Jain:2012qi}.

In a somewhat different development, a lower dimensional version of this duality 
was proposed in \cite{Gaberdiel:2010pz} and further refined in \cite{GG3}. 
It relates a one-parameter family of higher spin theories on
AdS$_3$ to a 't~Hooft like large $N$ limit of $2$-dimensional 
${\cal W}_{N,k}$ minimal model CFTs.\footnote{This is the natural
generalisation of the vector models in $3$ dimensions since, for 
vanishing 't~Hooft coupling, the theory is indeed equivalent to the singlet sector
of a free theory \cite{Gaberdiel:2011aa}.} This proposal was inspired by the asymptotic 
symmetry analysis (\`a la Brown-Henneaux \cite{Brown:1986nw})
of the higher spin theories \cite{Henneaux:2010xg,Campoleoni:2010zq} 
(see \cite{Gaberdiel:2011wb,Campoleoni:2011hg} for subsequent developments), and has,
by now, been tested in a variety of ways 
\cite{Gaberdiel:2011zw,Chang:2011mz,Papadodimas:2011pf,Ahn:2011by,Castro:2011iw,%
Ammon:2011ua,Chang:2011vk}. There have also been interesting results concerning the 
construction of  black holes for these higher spin theories, as well as their dual CFT 
interpretation  \cite{Gutperle:2011kf,Ammon:2011nk,Kraus:2011ds,GHJ}. 

The proposal of  \cite{Gaberdiel:2010pz} was generalised to the case 
where instead of the $\mathfrak{su}(N)$ based ${\cal W}$-algebras, one considers
the $\mathfrak{so}(2N)$ series \cite{Ahn:2011pv,Gaberdiel:2011nt}. More
recently, a ${\cal N}=2$ supersymmetric generalisation has been proposed
\cite{Creutzig:2011fe}, relating a family of Kazama-Suzuki models 
\cite{Kazama:1988qp,Kazama:1988uz} 
to the supersymmetric higher spin theory of  \cite{Prokushkin:1998bq,Prokushkin:1998vn},
and various aspects of it have been confirmed
\cite{Candu:2012jq,Henneaux:2012ny,Hanaki:2012yf,Ahn:2012fz}. 
\smallskip

As was already alluded to above, for the formulation of the original duality
\cite{Gaberdiel:2010pz} the determination of the asymptotic symmetry algebra
of the higher spin theory
\cite{Henneaux:2010xg,Campoleoni:2010zq,Gaberdiel:2011wb,Campoleoni:2011hg} was 
crucial since it defines, by the usual AdS/CFT correspondence, also the symmetry of
the dual CFT. Originally, this analysis was done classically, i.e.\  the algebra
was determined as a commutative Poisson algebra. However, since the resulting algebra
is non-linear, i.e.\ a  ${\cal W}$-algebra, the naive quantisation does not lead to a consistent
Lie algebra since normal-ordering contributions from commutators of non-linear terms spoil
the Jacobi identities. Recently, it was understood  \cite{GG3} how to overcome this
limitation for the original bosonic case. To this end the most general ${\cal W}_\infty$
algebra with the field content predicted by the asymptotic symmetry analysis
was studied. (For the bosonic case, the algebra is generated by one primary field for each
spin $s=2,3,4,\ldots$.) It was found that the Jacobi identities fix the structure of this algebra
up to two free parameters (see also \cite{CGKV}): the central charge $c$, as well as the
coupling constant $\gamma$ of the spin $s=4$ field in the 
OPE of two spin $s=3$ fields. This is what one would have expected for the 
quantisation of the classical asymptotic symmetry algebra since the latter has also
two free parameters: the size of AdS in Planck units --- this
is directly related to the central charge by the familiar Brown-Henneaux relation
\cite{Brown:1986nw} $c = \frac{3 \ell}{2 G}$ --- and the parameter characterising
the Lie algebra ${\rm hs}[\mu]$ whose Chern-Simons theory defines the higher spin theory.
In order to determine
the exact relation between $\gamma$ and $\mu$, the representation theory
of the two algebras was compared (see also \cite{Hornfeck:1993kp} for earlier work), using 
in particular that for $\mu=N$ integer, the
higher spin algebra  truncates to $\mathfrak{sl}(N)\cong {\rm hs}[N] / \chi_N$ for which the 
quantum representation theory is known; this then led to an explicit dictionary between
$\gamma$ and $\mu$, see eq.\ (2.15) of \cite{GG3}.

As it turned out, this relation is not one-to-one, meaning that there are different values 
of $\mu$ --- generically there are three distinct values ---  that lead to the {\em same} 
$\gamma$, and hence to the {\em same} ${\cal W}_{\infty}$ algebra; thus from the point of view 
of  ${\cal W}_{\infty}[\mu]$ there is a `triality' of identifications. One of these identifications
then implies the equivalence between the quantum symmetry of the bulk higher spin
theory, and the chiral algebra of the minimal model CFTs, and thus proves an 
important aspect of the conjectured duality of \cite{Gaberdiel:2010pz}. In fact,
the equivalence even holds for finite $N$ and $k$, and hence makes a prediction
for the `quantum corrections' of the higher spin theory that appear at finite $c$.

In this paper  we analyse the quantum ${\cal W}$ algebra of the ${\cal N}=2$ 
supersymmetric higher spin theory that was conjectured to be dual to
the 't~Hooft limit of the Kazama-Suzuki models in \cite{Creutzig:2011fe}. 
In this case, the higher spin algebra whose Chern-Simons action defines
the higher spin theory is ${\rm shs}[\mu]$, which
truncates, for $\mu=-N$, to the superalgebra $\mathfrak{sl}(N+1|N)$. 
The classical asymptotic symmetry algebra was already partially determined 
in \cite{Henneaux:2012ny,Hanaki:2012yf}, and first steps towards analysing
the quantum algebra were taken in \cite{Ahn:2012fz}, see also
\cite{blumenhagen2a} for earlier work. Here we follow the same
strategy as in \cite{GG3}: we first study the most general ${\cal N}=2$ supersymmetric
algebra $\sW_\infty$ whose field content agrees with that predicted by the asymptotic 
symmetry analysis of the higher spin theory (see section~2). Due to the complexity of 
the algebra, we can only study the first few commutators, but they already suggest
that the algebra $\sW_\infty$ is also characterised by two parameters: the central
charge $c$, and the coupling constant $\gamma$ of the Virasoro primary spin $s=2$ 
field $W^{2}$ in the OPE of $W^{2}$ with itself. 

In order to find the relation between
$\gamma$ and $\mu$ we then use (in section 3) that the Drinfel'd-Sokolov reduction of 
$\mathfrak{sl}(N+1|N)$ is equivalent to a Kazama-Suzuki model \cite{Ito:1990ac}.
The representation theory of the Kazama-Suzuki models follows directly from
their coset description, and thus again by comparing representations, we can
identify the exact relation between $\gamma$ and $\mu$, see eq.~(\ref{central}) below. 
We also check (see section~\ref{sec:wedge}) that the wedge subalgebra 
of $\sW_{\infty}[\mu]$ agrees indeed with $\shs[\mu]$, as has to 
be the case if the former is the Drinfel'd-Sokolov reduction of the latter \cite{Bowcock:1991zk}. 

As in the bosonic case of \cite{GG3}, the relation is not one-to-one, and there
are now generically four different values of $\mu$ that define the same algebra
(see section 4). 
Among other things this leads to the familiar level-rank duality of 
Kazama-Suzuki models 
that was already observed in \cite{Gepner:1988wi}. More importantly in our
context, the identifications also imply that 
the quantum  algebra $\sW_\infty[\lambda]$ of the higher spin gravity
theory is equivalent to the chiral algebra of the Kazama-Suzuki models. As before,
this relation does not just hold in the 't~Hooft limit, but even for finite $N$ and $k$; 
this establishes therefore an important aspect of the duality of
\cite{Creutzig:2011fe}. 

The structure of the quantum $\sW_\infty[\mu]$ algebra also implies how the
various representations behave as a function of $c$; for the case of the
two minimal representations that are dual to the scalar fields of \cite{Creutzig:2011fe}
this is studied in section~4.1. Following the same logic as in \cite{GG3} this analysis
suggests that in the present case 
both `scalar' fields should be thought of as describing non-perturbative
solutions in the semiclassical limit. We also suggest a possible explanation
of this somewhat surprising conclusion.

Finally, there are three appendices where we have collected some of the more technical 
material: appendix~A contains the commutation relations of the modes
of the low lying fields, while in appendix~B we give explicit expressions
for the first few composite fields. Finally, appendix~C describes the structure constants of the
superalgebra $\shs[\mu]$, as well as the relation between its generators and those of 
the wedge subalgebra of $\sW_{\infty}[\mu]$.


\section{The Structure of the ${\cal N}=2$ ${\cal W}$-algebras}


Let us begin by studying the structure of the superconformal ${\cal W}$-algebras 
that are relevant for the ${\cal N}=2$ version of minimal model holography
\cite{Gaberdiel:2010pz} proposed in \cite{Creutzig:2011fe}. These algebras, which
we shall denote by $\sW_\infty$ in the following, are generated, in addition to the 
${\cal N}=2$ superconformal algebra, by a single ${\cal N}=2$
primary field for every integer spin $s\geq 2$. The analysis of 
\cite{Creutzig:2011fe,Candu:2012jq,Henneaux:2012ny,Hanaki:2012yf,Ahn:2012fz} suggests
that, for each value of the central charge $c$, there is a one-parameter family
of such algebras that are labelled by the 't~Hooft parameter $\lambda$ of the Kazama-Suzuki
models. Here we want to show that, with the above field content, the Jacobi identities 
fix the algebra precisely up to two free parameters
that we can identify with the central charge $c$, and the self-coupling $\gamma$ 
of the spin-$2$ ${\cal N}=2$-primary field. In the next section we shall 
then explain how $\gamma$ can be expressed in terms of $\lambda$ and $c$. 

Recall that each ${\cal N}=2$ multiplet contains 4 Virasoro primary fields. Indeed, if we 
denote by $W^{s}$ the ${\cal N}=2$ primary field of spin $s$ and $\U(1)$-charge zero, 
then the four fields are simply
\be\label{Virpri}
W^{s\, 0} = W^{s} \ , \qquad
W^{s\, \pm} = G^\pm_{-\frac{1}{2}} W^{s} \ , \qquad
W^{s\, 1} = \tfrac{1}{4} \bigl(G^+_{-\frac{1}{2}}\, G^-_{-\frac{1}{2}} -  
G^-_{-\frac{1}{2}}\, G^+_{-\frac{1}{2}} \bigr) W^{s} \ .
\ee
Here we have used the usual conventions for the ${\cal N}=2$ superconformal algebra
which we review for the convenience of the reader in appendix~A. The fields
$W^{s\, \pm}$ have spin $s+\frac{1}{2}$ and $\U(1)$-charge $\pm 1$, while the field
$W^{s\, 1}$ has spin $s+1$ and $\U(1)$-charge zero. The fact that they lie in an ${\cal N}=2$
multiplet means that the various components satisfy the OPEs
{
\allowdisplaybreaks
\begin{align}\label{eq:primary2opes}
G^\pm(z) W^{s\, 0}(w)&\sim \mp \frac{W^{s\, \pm}(w)}{z-w}\ ,\qquad \qquad
G^\pm(z) W^{s\, \pm} (w)\sim 0\ ,\\ \notag
G^\pm(z) W^{s\, \mp} (w)&\sim \pm 
\left[\frac{ 2s W^{s\, 0}(w)}{(z-w)^2}+\frac{\partial W^{s\, 0}(w)}{z-w}\right]
+\frac{2W^{s\, 1}(w)}{z-w}\ ,\\ \notag
G^\pm(z)W^{s\, 1}(w)&\sim \frac{1}{2}\left[\frac{(2s+1)W^{s\, \pm}(w)}{(z-w)^2}+
\frac{\partial W^{s\, {\pm}}(w)}{z-w}\right]\ ,\\ \notag
J(z) W^{s\, 1}(w)&\sim s\, \frac{ W^{s\, 0}(w)}{(z-w)^2} \ .
\end{align}
}
Using the usual expansion of fields in terms of modes
\be
W(z) = \sum_{n\in\mathbb{Z}} W_n z^{-n-h} \ , 
\ee
where $h$ is the conformal dimension of $W$, the OPEs~\eqref{eq:primary2opes} 
can also be converted into
commutation relations for the corresponding modes; the resulting formulae are given 
in (\ref{WsNcom}).

We shall collectively denote the fields in the $\mathcal{N}=2$ multiplet~\eqref{Virpri} 
by $W^{(s)}$; in order to have a coherent notation, we shall also denote
the generating fields $J$, $G^{\pm}$, and  $T$ of the 
$\mathcal{N}=2$ Virasoro algebra by $W^{(1)}$.

%


\subsection{The Strategy}
\label{sec:boot}



Next we want to study the OPEs of the fields $W^{s\, \alpha}$, $\alpha=0,\pm,1$ 
with one another. These
OPEs are constrained by the requirement that they must be associative; translated into
modes this is believed to be equivalent to the condition that the corresponding commutators
satisfy the Jacobi identity.

We shall proceed in two steps. First we present the most general ansatz for the 
singular part of the OPEs $W^{s_1\, \alpha_1}(z) \,  W^{s_2\, \alpha_2}(w)$ with
$s_1,s_2\geq 2$ that is compatible with the full ${\cal N}=2$ superconformal symmetry
and with the assumed spectrum of $\sW_\infty$.
This step is actually the technical core of our calculation: we have worked in terms of
Virasoro primaries, using the 
{\em Mathematica} packages {\tt OPEdefs} and {\tt OPEconf} of Thielemans\footnote{The 
latest versions of {\tt OPEdefs} and {\tt OPEconf} are available directly from the author.} 
\cite{opedefs, thelemansthesis}. The compatibility with the ${\cal N}=2$ superconformal 
symmetry can then be implemented by requiring the associativity of the OPE with the 
${\cal N}=2$ superconformal generators. This fixes the coefficients of the various 
Virasoro primaries relative to one another.

In a second step we then require that these different OPEs are associative, i.e.\  that they
satisfy
\begin{equation}\label{eq:assoc}
\Bigl( \Bigl( W^{s_1\, \alpha_1}(x) W^{s_2\, \alpha_2}(y)\Bigr) 
W^{s_2\, \alpha_3}(z)\Bigr) =
 \Bigl( W^{s_1\, \alpha_1}(x)  \Bigl( W^{s_2\, \alpha_2}(y)
W^{s_2\, \alpha_3}(z)\Bigr) \Bigr) \ . 
\end{equation}
It is believed that requiring (\ref{eq:assoc}) is
equivalent to demanding the Jacobi identities
\be\label{Jacobigen}
\Bigl[\, W^{s_1\, \alpha_1}_{m_1} , \Bigl[ W^{s_2\, \alpha_2}_{m_2}, W^{s_3\, \alpha_3}_{m_3}
\Bigr] \, \Bigr] 
+ \hbox{cycl.} = 0 \qquad \hbox{for all $m_1,m_2,m_3$.}
\ee
Note that in order to study (\ref{eq:assoc}), one has to work with the \emph{full} OPEs,
rather than just their singular part. 
The precise way in which this calculation can be done is explained in detail 
in the thesis of Thielemans \cite{thelemansthesis}. His  
\emph{Mathematica} package {\tt OPEdefs} allows to compute these
associativity constraints very efficiently by using a built-in function called {\tt OPEJacobi}.

We shall present our answers mostly in a rather compact form, 
namely by grouping together the fields that appear in the same ${\cal N}=2$ 
superconformal representation. However, as mentioned before, 
we have actually carried out the calculations by working in terms of Virasoro primaries
and then realising the $\mathcal{N}=2$ superconformal symmetry
by solving the $W^{(1)}\times W^{(s_1)}\times W^{(s_2)}$ associativity constraints.

For the case at hand, there are infinitely many ${\cal N}=2$ superconformal primary fields,
and therefore infinitely many associativity constraints to check. The full problem is therefore
too hard to be solved completely. However, we have studied the low-lying OPEs in detail, and 
they already suggest that there is indeed exactly one free parameter beyond the central charge,
that characterises these $\sW_\infty$ algebras.

\subsection{Enumerating ${\cal N}=2$ Primary Fields}
\label{sec:ansatz_alg}


Before we can make the most general ansatz for the various OPEs, we first need
to understand how many ${\cal N}=2$ primary fields $\sW_\infty$ contains. (In particular,
we need to determine how many composite ${\cal N}=2$ primary fields there are.)
This information can be easily read off from the vacuum character\footnote{In the definition
of the various characters we drop for convenience the overall $-\frac{c}{24}$ exponent.}
 of $\sW_\infty$
\begin{equation}
\chi_\infty(q,z) = \hbox{Tr}_0 \bigl( q^{L_0} z^{J_0} \bigr) = 
\prod_{s=1}^\infty \prod_{n=s}^\infty
\frac{(1+z \, q^{n+\frac{1}{2}})(1+z^{-1} q^{n+\frac{1}{2}})}{(1-q^n)(1-q^{n+1})} \ .
\end{equation}
We want to decompose $\chi_\infty(q,z)$ in terms of characters of irreducible
$\mathcal{N}=2$ Virasoro representations. The character of the ${\cal N}=2$
vacuum representation equals 
\be\label{N2vac}
\chi_{0} (q,z) = \prod_{n=1}^\infty
\frac{(1+z\,  q^{n+\frac{1}{2}})(1+z^{-1} q^{n+\frac{1}{2}})}{(1-q^n)(1-q^{n+1})}  \ ,
\ee
while for a generic irreducible ${\cal N}=2$ representation with highest weight $(h,Q)$ 
with respect to $(L_0,J_0)$ we have instead 
\begin{align}
\chi_{(h,Q)}(q,z) & = q^h \, z^Q\, \prod_{n=1}^\infty
\frac{(1+z\, q^{n-\frac{1}{2}})(1+z^{-1} q^{n-\frac{1}{2}})}{(1-q^n)^2}  \notag \\
& = q^h\, z^Q\, \frac {(1+z\, q^{\frac{1}{2}}) ( 1+z^{-1}q^{\frac{1}{2}})}{(1-q)}\, \chi_{0} (q,z)  \ .
\end{align}
The multiplicity $d(h,Q)$  of an ${\cal N}=2$ primary field with quantum numbers 
$(h,Q)$ in $\sW_\infty$  is then simply determined by the decomposition
\be
\chi_\infty(q,z) = \chi_{0} (q,z) + \sum_{h\in \frac{1}{2}\mathbb{N}}\, 
\sum_{Q\in\mathbb{Z}} \, d(h,Q)\, \chi_{(h,Q)}(q,z) \ .
\ee
Writing $\chi_\infty(q,z) = \chi_0(q,z) \cdot \chi_{\rm HS}(q,z)$ with 
\be\label{HSdef}
\chi_{\textrm{HS}}(q,z)=\prod_{s=2}^\infty \prod_{n=s}^\infty
\frac{(1+ z q^{n+\frac{1}{2}})(1+z^{-1} q^{n+\frac{1}{2}})}{(1-q^n)(1- q^{n+1})} \ ,
\ee
and diving the whole expression by $\chi_0(q,z)$, 
the generating function for $d(h,Q)$ turns out to equal
\be
P(q,z)\equiv 
\sum_{h\in\frac{1}{2}\mathbb{N}}\, \sum_{Q\in \mathbb{Z}}
d(h,Q) \, q^h\, z^Q = 
\frac{(1-q)\bigl(\chi_{\textrm{HS}}(q,z)-1\bigr)}{(1+z\, q^{\frac{1}{2}})(1+z^{-1}q^{\frac{1}{2}})} \ .
\ee
The first few terms are explicitly
\be\label{eq:p2_series}
P(q,z) = q^2 + q^3 + 2 q^4 + 3 q^5 + (2  z + 2 z^{-1} ) q^{\frac{11}{2}}
+ 7 q^6 + \cdots \ . 
\ee
Since in $\sW_{\infty}$ 
there is one simple  ${\cal N}=2$ primary field for every spin  $s\geq 2$, we read off
from (\ref{eq:p2_series}) that the first composite ${\cal N}=2$ primary field appears at spin $4$
and $\U(1)$ charge zero; this field is essentially the normal ordered product of $W^{(2)}$ with
itself. The higher terms can be similarly interpreted. 

\subsection{Constraining the OPE}

Let us illustrate our method with the example of the OPEs of the ${\cal N}=2$ 
supermultiplet $W^{(2)}$ (whose ${\cal N}=2$ primary is the spin $2$ field $W^2$). 
In terms of ${\cal N}=2$ multiplets, the singular part of the 
OPE has the general form
\begin{equation}\label{eq:ope22}
W^{(2)}\times W^{(2)} \sim n_2 I + c_{22,2} \, W^{(2)} + c_{22,3} \, W^{(3)} \ ,
\end{equation}
where on the right-hand-side also the corresponding ${\cal N}=2$
superconformal descendants are included (if they contribute to the singular part of the
OPE).

In order to see that this is the most general ansatz recall that in the singular part 
of the OPE of two Virasoro primary
fields of conformal dimension $h_1$ and $h_2$, 
only Virasoro primary fields with $h\leq h_1 + h_2 - 1$ can appear. 
However, to apply this general rule to our current context,
we need to remember that each ${\cal N}=2$ multiplet actually contains $4$ Virasoro
primaries, see eq.~(\ref{Virpri}). Thus each ${\cal N}=2$ OPE gives actually rise to $16$
OPEs of Virasoro primaries; the condition that an  ${\cal N}=2$ multiplet
appears in the OPE then requires that all its $4$ Virasoro primaries  of eq.~(\ref{Virpri})
appear among the $16$ Virasoro primary OPEs.\footnote{Another way of saying this
is that the ${\cal N}=2$ primary of the right-hand-side does not necessarily have to 
appear in the OPE of the two ${\cal N}=2$ primaries on the left-hand-side. Indeed,
this is the origin of the so-called odd fusion rules of \cite{Mussardo:1988av}, see
also \cite{Gaberdiel:1993mt}.} Obviously, the ${\cal N}=2$ superconformal
symmetry relates the structure constants of some of these Virasoro primaries to one another,
but the explicit expressions are somewhat complicated, see \cite{blumenhagen2}.

Given the structure of (\ref{eq:p2_series}) it follows that the singular part of the OPE can, apart
from the identity $I$ of spin zero, at most contain the ${\cal N}=2$ multiplets $W^{(s)}$ of spin 
$s=2,3,4$. The $\mathcal{N}=2$ multiplet  of spin $s=4$, however, cannot actually appear,
since it contains the Virasoro primary $W^{4\, 1}$ of spin $s=5$. However, 
$W^{4\, 1}$ can only appear in the OPE $W^{2\, 1} \times W^{2\, 1}$, and then the conformal
symmetry requires that 
the coefficient of $W^{4\, 1}_{m+n}$  in the commutator
$[W^{2\, 1}_m,W^{2\, 1}_n]$ is independent of $m$ and $n$. Since the commutator
must be anti-symmetric in $m\leftrightarrow n$, the overall coefficient must therefore vanish. 
Thus we arrive at (\ref{eq:ope22}). 


We also need to be specific about what we precisely mean by the various structure constants,
given that the ${\cal N}=2$ primary does not necessarily appear
in the OPE of the two ${\cal N}=2$ primaries.
%
%
%
%
%
%
We define $c_{22,2}$ by the OPE
\begin{align}\label{eq:structconst}
W^{2\, 0}(z) W^{2\, 0}(w) \sim & \frac{n_2}{(z-w)^4}
+ \frac{n_2}{c(c-1)}\, \frac{(4c T -6 J^2)(w)}{(z-w)^2}
+ c_{22,2}\, \frac{W^{2\, 0}(w)}{(z-w)^2}\\ \notag
{}&+
\frac{n_2}{c(c-1)}\, \frac{(2c \partial T-6\partial JJ)(w)}{(z-w)}+
\frac{c_{22,2}}{2}\, \frac{\partial W^{0\, 2}(w)}{z-w}\ ,
\end{align}
where we have for once given {\em all} the singular terms.
(In the following we shall not be so explicit any more, 
see however appendix~\ref{sec:22com} for the commutators of $W^{2\, \alpha}$.)
The structure constant $c_{22,3}$ can be similarly defined by a coefficient in the OPE
\begin{align}\label{eq:structconst1}
W^{2\, 0}(z) W^{2\, -}(w) &\sim 
\frac{c_{22,2}}{2}\, \frac{W^{2\, -}(w)}{(z-w)^2} + c_{22,3} \,\frac{W^{3\, -}(w)}{z-w}+\cdots \ .
\end{align}

So far we have only used the constraints that come from the ${\cal N}=2$ superconformal
symmetry, i.e.\ the conditions that follow from the Jacobi identities (\ref{Jacobigen})
associated to $s_1=1$ and $s_2=s_3=2$. The next step is to study (\ref{Jacobigen})
(or the associativity of (\ref{eq:assoc})) for $s_1=s_2=s_3=2$. However, since 
the OPE of $W^{(2)}$ with itself generates $W^{(3)}$, we also need to make an ansatz for 
$W^{(2)} \times W^{(3)}$. Using the same arguments as above, one finds that the 
most general ansatz for the singular part of that OPE is (again the ${\cal N}=2$
superconformal descendants are included where they contribute to the singular part)
\begin{align}\label{eq:ope23}
W^{(2)}\times W^{(3)} & \sim 
c_{23,2}\,  W^{(2)} + c_{23,3}\, W^{(3)} + c_{23,4}\,  W^{(4)} 
+ a_{23,4} A^{(4)}\\
{}&\quad + c_{23,5} W^{(5)} + a_{23,5} A^{(5)} \ ,
\notag
\end{align}
where $A^4$ and $A^5$ are the uncharged composite ${\cal N}=2$ primaries of 
spin 4 and 5, respectively. Note that it would seem from (\ref{eq:p2_series}) that 
three fields of spin $5$ should generically appear on the right-hand-side of (\ref{eq:ope23}). However,
one of them is just the normal ordered product of $W^2$ with $W^3$ (appropriately
completed to make it ${\cal N}=2$ primary), which does not contribute to the singular
part of the OPE.
The composite fields $A^4$ and $A^5$  have the leading terms
%
\begin{align*}
A^4 &= \left(W^{2\, 0}\right)^2+\cdots\ ,\\
A^5 &= W^{2\, 0}W^{2\, 1}+
\tfrac{(5c-24) }{8 c} \, W^{2\, +}W^{2\, -}-
\tfrac{6 }{c} \, J\left(W^{2\, 0}\right)^2+\cdots
\end{align*}
that have to be completed to make them $\mathcal{N}=2$ primaries; explicit
formulae for them are given in appendix~\ref{sec:composite}.
The structure constants in the ansatz~\eqref{eq:ope23} are again defined by 
OPE coefficients as
\begin{align} \label{eq:str3}
W^{2\, +}(z) W^{3\, -}(w) \sim\ & c_{23,2}\,\frac{20 W^{2\, 0}(w)}{(z-w)^4}+
c_{23,3}\, \frac{2W^{3\, 0}(w)}{(z-w)^3}-c_{23,4}\, \frac{8 W^{4\, 0}(w)}{(z-w)^2}\\
{}&-a_{23,4}\, \frac{8 A^{4\, 0}(w)}{(z-w)^2}
+
c_{23,5} \, \frac{W^{5\, 0}(w)}{z-w}
+ a_{23,5}\, \frac{ A^{5\, 0}(w)}{z-w}+\cdots \ .
\notag
\end{align}
It follows from 
the permutation symmetry of the $3$-point functions together with the
${\cal N}=2$ superconformal symmetry, see the appendix of \cite{blumenhagen2},  that we have the relation 
\begin{equation}\label{eq:223vs232}
10 \, c_{23,2}\,  n_2 = - 3 \, c_{22,3} \, n_3\ ,
\end{equation}
where $n_s$ are the normalisation constants
\begin{equation}\label{eq:norm}
\langle W^{s\, 0}(z) W^{s\, 0}(w)\rangle = \frac{n_s}{(z-w)^{2s}}\ .
\end{equation}
This fixes the normalisation of the other fields in the multiplet as well.


\subsection{Structure Constants}


Now we have everything in place to study the associativity of the OPE (\ref{eq:assoc})
for $s_1=s_2=s_3=2$. The calculation is somewhat tedious, but the end result is simple:
the structure constants appearing in (\ref{eq:ope22}) and (\ref{eq:ope23}) must
satisfy 
\begin{align}\label{eq:soljac1}
c_{22,3} \, c_{23,2} &= -\frac{6(c+3)(5c-12)}{5c(c+6)(2c-3)}\, n_2
-\frac{3(c-15)(c-1)}{10(c+3)(5c-12)}\, \left(c_{22,2}\right)^2\\ \label{eq:soljac2}
c_{23,3} &= \frac{3(c+6)(2c-3)}{(c+3)(5c-12)}\, c_{22,2} \\ 
c_{23,5} &= a_{23,5} = 0\ . \label{eq:soljac3}
\end{align}
Before proceeding we should note that 
there is one non-trivial consistency check we can easily perform: if we require all 
$W^{(s)}$ multiplets with $s\geq 3$ to vanish, our algebra should reduce to that 
constructed explicitly by Romans in \cite{romans}. In order to be able to decouple
all of these fields, we need in particular that $c_{22,3}=0$ in eq.~\eqref{eq:ope22}.
Eq.~\eqref{eq:soljac1} then fixes $c_{22,2}$ in terms of $n_2$, and the solution 
coincides exactly with the one obtained in \cite{romans} if we normalise the 
$W^{2\, 0}$ field as he does, namely with $n_2=\frac{c}{2}$.
\smallskip

Next we observe that eqs.~(\ref{eq:223vs232}) and (\ref{eq:soljac1}) -- (\ref{eq:soljac3}) 
fix all the structure constants in the OPEs (\ref{eq:ope22}) and (\ref{eq:ope23}), 
except for $c_{22,2}$, $c_{23,4}$ and $a_{23,4}$. The fact that the latter two 
structure constants are unconstrained at this stage is not surprising: it simply reflects
that we can rescale $W^4$ and $A^4$ arbitrarily. (Furthermore, there is the freedom
of redefining $W^4$ by adding to it a multiple of $A^4$ --- as a consequence, these
two structure constants must always appear together in Jacobi identities \cite{KV}.)
Thus there is only one free parameter at this stage, namely the structure
constant $c_{22,2}$. Note that our normalisation convention (\ref{eq:norm}) only
fixes $W^2$ up to a sign, leading to a sign ambiguity in the definition of $c_{22,2}$.

Thus the structure seems to be rather similar to that of the non-supersymmetric
${\cal W}$-algebra ${\cal W}_\infty[\mu]$, that is determined (for each central
charge $c$) by a single structure constant \cite{GG3}, see also \cite{CGKV}.
Reasoning by analogy with \cite{GG3} we therefore conjecture that 
for every value of the central charge $c$, there exists  a one-parameter
family of non-isomorphic $\sW_{\infty}$ algebras which are parametrised by 
\begin{equation}\label{eq:par}
\gamma =\left(c_{22,2}\right)^2\ .
\end{equation}
In the next section we want to relate these algebras to those that appear in the 
supersymmetric minimal model holography of \cite{Creutzig:2011fe}.


\section{Minimal Representation}


In the application to minimal model holography \cite{Gaberdiel:2010pz,Creutzig:2011fe}, 
the above $\sW_\infty$ algebras should arise as the Drinfel'd-Sokolov (DS) reduction of
the infinite dimensional Lie algebra ${\rm shs}[\mu]$, which can be constructed as 
\be\label{shsdef}
{\rm shs}[\mu] \oplus \mathbb{C} = \frac{U(\mathfrak{osp}(1|2))}{\langle C^{\mathfrak{osp}} 
- \frac{1}{4} \mu (\mu-1) {\bf 1} \rangle} \ .
\ee
Here  we have normalised the Casimir operator $C^{\mathfrak{osp}}$ so that 
for the $\mathfrak{osp}(1|2)$ representation of dimension $4j+1$ it takes the value 
$C^{\mathfrak{osp}}=j(j+\frac{1}{2})$. We can think of ${\rm shs}[\mu]$ as 
\be
{\rm shs}[\mu] \cong \mathfrak{sl}(1-\mu | -\mu) 
\ee
since, for $\mu=-N$ with $N\in\mathbb{N}$, we have\footnote{Note that
by definition ${\rm shs}[-\mu] \cong {\rm shs}[1+\mu]$.}
\be
{\rm shs}[-N]  / \chi_N \cong \mathfrak{sl}(N+1 | N)  \ .
\ee
Here $\chi_N$ is the maximal (infinite-dimensional) ideal  that appears for these values of $\mu$. Let
us denote the DS-reduction of ${\rm shs}[\mu]$ by $\sW_\infty[\mu]$,
\be
\sW_\infty[\mu] \equiv \hbox{Drinfel'd-Sokolov reduction of}\  {\rm shs}[\mu] \ .
\ee
We want to understand how to relate $\gamma$ in (\ref{eq:par}) to $\mu$.

It was shown by Ito in \cite{Ito:1990ac} that the Drinfel'd-Sokolov reduction of 
$\mathfrak{sl}(N+1 | N)$ is equivalent to the Kazama-Suzuki coset 
\begin{equation}\label{eq:kzcoset}
\frac{\SU(N+1)_k\times \SO(2N)_1}{\SU(N)_{k+1}\times \U(1)_\kappa}\ ,
\end{equation}
where $\kappa = N(N+1)(N+k+1)$, see \cite{Candu:2012jq} for our notation. 
Thus it follows that (\ref{eq:kzcoset}) is equivalent to (a quotient of) $\sW_\infty[-N]$. For future
use we also recall that the central charge of (\ref{eq:kzcoset}) equals
\be\label{cNk}
c_{N,k} = \frac{3kN}{k+N+1} \ .
\ee
\medskip

In order to relate $\gamma$ to $\mu$ we can now use the same idea as 
in \cite{Hornfeck:1993kp,GG3}. From their coset description, it is clear that the 
Kazama-Suzuki models possess {\em minimal} representations whose 
Virasoro character agrees to low orders in $q$ with 
\begin{equation}\label{eq:oldchmin}
\chi_{\text{min}}(q) = \frac{q^h (1+q^{\frac{1}{2}})}{(1-q)}\, 
\prod_{s=1}^{\infty}\prod_{n=s}^{\infty} \frac{(1 + q^{n+\frac{1}{2}})^2}{(1-q^n)(1-q^{n+1})} \ . 
\end{equation}
Indeed, in the notation of \cite{Candu:2012jq} where the coset representations are labelled by 
$(\Lambda;\Xi,l)$, this is the case for the $4$ representations
\be\label{minimal}
 ({\rm f};0,N) \ , \qquad (\bar{\rm f};0,-N) \ , \qquad
(0;{\rm f},-(N+1)) \ , \qquad (0;\bar{\rm f},(N+1) ) \ ,
\ee
where ${\rm f}$ and $\bar{\rm f}$ is the fundamental and anti-fundamental representation of 
${\rm SU}(N)$ or ${\rm SU}(N+1)$, respectively. Their conformal dimensions equal
\be\label{hNk}
\begin{array}{l}
{\displaystyle 
h\bigl( ({\rm f};0,N)  \bigr)  = h\bigl( (\bar{\rm f};0,-N)  \bigr) =  \frac{N}{2(N+k+1)}}  \vspace*{0.2cm} \\
{\displaystyle 
h\bigl( (0;{\rm f},-(N+1))  \bigr)  = h\bigl( (0;\bar{\rm f},N+1)  \bigr) = \frac{k}{2(N+k+1)} \ , }\end{array}
\ee
and all of them are `chiral primaries', i.e.\ have $Q=\pm  2 h$. Indeed, this is immediate
from their character formula (\ref{eq:oldchmin}), which implies that each of
these representations has a null-vector of the form $G^\pm_{-1/2} |h,Q\rangle$. 
Furthermore, since the character has only a single state at conformal weight $h+\tfrac{1}{2}$,
all other $(-1/2)$-descendants have to be proportional to $G^-_{-1/2} P^0$, i.e.\
the representation generated from $P^0$ has to have very many null-vectors. 

These null-vectors are only compatible with the commutation relations of $\sW_{\infty}$
(that depend on $\gamma$) if $h$ solves an equation in terms of $\gamma$ and $c$. 
We can then compare this to the solutions (\ref{hNk}) that arise for $\mu=-N$
and $c=c_{N,k}$, and this will allow us to determine the $N$ (and $c$) dependence of 
$\gamma$; analytically continuing in $N$ will then finally lead to the desired relation
between $\mu$ and $\gamma$. 


\subsection{Ansatz for OPEs}


Actually, it will be more convenient to work out the equation for $h$ in terms of $\gamma$ 
not directly using the commutation relations (as was done in the bosonic case in \cite{GG3}), 
but rather by exploiting the associativity of the OPE $W^{(2)} \times W^{(2)} \times P$.
(Incidentally, this is also the approach that was taken in the original analysis of 
\cite{Hornfeck:1993kp}.) In order to do so, we first need to translate the above statements
about the structure of the $P$-representation, into the OPE language.
For definiteness, let us consider the case $Q=+2h$, so that $G^+_{-1/2} P^{0}=0$,
where $P^{0} = |h,Q\rangle$ is the ${\cal N}=2$ primary.  (The case  $Q=-2h$ works
analogously.) Let us also denote the non-vanishing $G^-_{-1/2}$ descendant 
of conformal weight $h+\tfrac{1}{2}$ by $P^-$. 
The above statements about the null-vectors then imply that we have the OPEs 
\begin{align}\label{eq:chiralope}
G^-(z) P^{0}(w) &\sim \frac{P^{-}}{z-w}\ ,& G^+(z) P^{0}(w) &\sim 0\ ,\\
G^+(z)P^{-}(w)&\sim \frac{4h P^{0}}{(z-w)^2}
+\frac{2\partial P^{0}}{z-w}\ , & G^-(z) P^{-}(w) & \sim 0\ .
\end{align}
%

In order to work out the most general ansatz for the OPEs of the higher spin fields $W^{(s)}$ with 
$P$, we need to understand again the decomposition of the minimal representation in terms of
irreducible ${\cal N}=2$ representations. As before in section~\ref{sec:ansatz_alg}, this can
be read off from the character (\ref{eq:oldchmin}). Including the $\U(1)$-chemical 
potential $z$, the character of the minimal representation (with $Q=2h$) has the form
\begin{align}
\chi_{\text{min}}(q,z) & = q^h z^{2h} \frac{(1+z^{-1} q^{\frac{1}{2}} )}{(1-q)} \, 
\prod_{s=1}^{\infty} \prod_{n=s}^{\infty} \frac{(1+z q^{n+\frac{1}{2}}) (1+z^{-1} q^{n+\frac{1}{2}}) }
{(1-q^n) (1-q^{n+1})} \\
& = q^{h} z^{2h} \, \chi_0 \Bigl[ 
 \frac{(1+z^{-1} q^{\frac{1}{2}} )}{(1-q)}
+ \sum_{s\in\frac{1}{2}\mathbb{N}} \sum_{Q\in\mathbb{Z}} d_{\text{min}}(s,Q) \, q^{s} z^Q\, 
\frac{(1+z q^{\frac{1}{2}})(1+z^{-1} q^{\frac{1}{2}})}{(1-q)} \Bigr] \ , \notag
\end{align}
where $\chi_0$ was defined in (\ref{N2vac}), and 
$d_{\text{min}}(s,Q)$ is the multiplicity of the ${\cal N}=2$ representation
with conformal dimension $h'=h+s$ and $\U(1)$-eigenvalue $Q'=2h+Q$. 
Their generating function is now
\begin{equation}
P_{\mathrm{min}}(q,z) =  \sum_{s\in\frac{1}{2}\mathbb{N}}\sum_{Q\in\mathbb{Z}} 
d_{\text{min}}(s,Q) \, q^{s} z^Q = 
\frac{\chi_{\textrm{HS}}(q,z) - 1}{1+zq^{\frac{1}{2}}} \ ,
\end{equation}
where $\chi_{\textrm{HS}}$ was defined in (\ref{HSdef}). 
The first few terms are explicitly
\begin{equation}\label{eq:minpct}
P_{\mathrm{min}}(q,z) = q^2+ z^{-1} q^{\frac{5}{2}} +2\,  q^3 + 2\, z^{-1} q^{\frac{7}{2}}+\cdots\ .
\end{equation}
With these preparations, we can now make the most general ansatz for the singular part of the 
OPE of  $P$ with the higher spin fields 
\begin{align}\label{eq:ope2}
W^{(2)}\times P &\sim w_2 \, P \ , \\
W^{(3)}\times P & \sim w_3\, P + a\, P^{(2)} + b\, P^{(\frac{5}{2})}\ ,
\label{eq:ope3}
\end{align}
where $P^{(2)}$ and $P^{(\frac{5}{2})}$ are the composite $\mathcal{N}=2$ (non-chiral) 
primary fields corresponding to the first two terms in eq.~\eqref{eq:minpct}; their
corresponding ${\cal N}=2$ primary states are of the form
\begin{equation}
P^{2\, 0} = W^{2 \, 0}_{-2} \, P^{0}+\cdots\ , \qquad 
P^{\frac{5}{2} \, 0} =  W^{2 \, 0}_{-2} \, P^{-} - h W^{2\, -}_{-\frac{5}{2}}P^0 + \cdots \ , 
\end{equation}
completed to make them ${\cal N}=2$ primary, see appendix~B for the full expressions.
The parameters appearing in (\ref{eq:ope2}) and (\ref{eq:ope3})
are defined in terms of OPE coefficients of the component fields as 
\begin{align}
 W^{2\,0 }(z)\,  P^{0}(w) & \sim w_2 \left[ \frac{P^{0}(w)}{(z-w)^2}
+\frac{1}{h(c-6h)}\, \frac{c\partial P^0(w)-6h (JP^0)(w)}{z-w}
\right] \\
 W^{3\, 0}(z)\, P^{0}(w) &\sim w_3 \, \frac{P^{0}(w)}{(z-w)^3}+
 a\, \frac{P^{2\, 0}(w)}{z-w}+\cdots \\
W^{3\, 0}(z)\, P^{-}(w) &\sim b\, \frac{P^{\frac{5}{2}\, 0}(w)}{(z-w)}+\cdots \ .
\end{align}
In particular, $w_2$ and $w_3$ are therefore the eigenvalue of $P^0$ with
respect to $W^{2\, 0}_0$ and $W^{3\, 0}_0$, respectively.


\subsection{Structure Constants}


Now we are in the position to study the associativity of the OPEs 
$W^{(2)}\times W^{(2)}\times P$. After a tedious but straightforward calculation it 
leads to the constraints on the parameters appearing
in eqs.~(\ref{eq:ope2}) and (\ref{eq:ope3}) 
\begin{align}
w_2 ^2 &= \frac{ h^2(1+2h)(c-6h)(c-3+12h) n_2 }{ c(c-1)[c(1-h)+3h]} \label{eq:w2} \\
w_3 & = -\frac{ 12h(1+h)(c-1)(c+3)(c-12h)(c-6+18h) n_2}{ c(c+6)(2c-3)(5c-12)[c(1-h)+3h] 
c_{22,3}}\\
a & = \frac{72(1+h)(c-1)(c-6+18h) w_2}{(1+2h)(5c-12)(c-6h)(c-3+12h) c_{22,3}}\\
b & = -\frac{ 54(2h-1)(c-1)(c-12h) w_2}{ h(1+2h)(5c-12)(c-6h)(c-3+12h) c_{22,3}} \ .
\end{align}
Furthermore, the conformal dimension $h$ of the minimal representation must be 
related to the parameter $c_{22,2}$ of the $\sW_{\infty}$ algebra by
\begin{equation}\label{eq:hresult}
 c_{22,2} = -\frac{2(c+3)[c(1-4h)-12h^2]w_2}{h(1+2h)(c-6h)(c-3+12h)}\ .
\end{equation}
%
Note that it follows that the structure constants of eqs.~(\ref{eq:ope2}) and (\ref{eq:ope3})
are uniquely determined (up to the sign of $w_2$) by the associativity of the OPEs
$W^{(2)} \times W^{(2)}\times P$.
In terms of the parameter~\eqref{eq:par} and using eq.~(\ref{eq:w2}) for $w_2^2$, 
the relation~\eqref{eq:hresult} then finally becomes
\begin{equation}\label{eq:gammahc}
\gamma = \frac{ 4(c+3)^2[c(1-4h)-12h^2]^2n_2 }{ (1+2h)c(c-1)(c-6h)(c-3+12h)[c(1-h)+3h] }\ .
\end{equation}

\subsection{The Desired Relation}

Next we plug into eq.~\eqref{eq:gammahc} the expression for the central charge 
$c=c_{N,k}$ \eqref{cNk}, and  one of the conformal dimensions $h$ in eq.~\eqref{hNk};
this leads to a relation for $\gamma$ in terms of $N$ and $k$
\begin{equation}\label{eq:gammahk}
\gamma = \frac{8 (1 + k)^2 (k - N)^2 (1 + N)^2 (1 + k + N) n_2}{( k-1) k ( N-1) 
N (1 + 2 k + N) (1 + k + 2 N) (3 k N-N-k-1)}\ .
\end{equation}
Note that the same formula is obtained, independent of which of the two solutions in (\ref{hNk})
one considers; this is a non-trivial consistency check on our analysis. 
Next we want to replace
$k$ in favour of $c$; unlike the bosonic case considered in \cite{GG3}, here 
the central charge~\eqref{cNk} uniquely determines the level $k$, and we get 
\begin{equation}\label{eq:ktoc}
 k = \frac{c \, (N+1)}{3N-c}\ .
\end{equation}
%
%
%
Plugging this relation into eq.~\eqref{eq:gammahk} we then get an expression for 
$\gamma$ as a function of $N$ and $c$ 
\begin{equation}\label{eq:gammacN}
\gamma = -\frac{8 (c+3)^2 (c + 2 c N - 3 N^2)^2 n_2}{c(c-1) ( c -3- 6 N) (N-1) (c + 3 N) (2 c - 3 N + 
    c N)}\ .
\end{equation}
Finally, we can replace $N$ by $-\mu$ in the above equation, and analytically continue 
$\mu$; this leads to our central relation
\be\label{central}
\boxed{ \gamma(\mu,c) = \frac{8 (c+3)^2 \, (c - 2 c \mu - 3 \mu^2)^2 \, n_2}{c(c-1) 
( c -3+ 6 \mu) (\mu+1) (c - 3 \mu) (2 c + 3 \mu -  c \mu)}}
\ee
establishing the connection between the  $\gamma$-parameter of the $\sW_\infty$ algebra, 
and the $\mu$-parameter in $\sW_{\infty}[\mu]$, i.e.\ in the DS reduction 
of $\mathrm{shs}[\mu]$. Note that the $n_2$ factor on the right-hand-side simply reflects
the fact that $\gamma=(c_{22,2})^2$ depends on the normalisation of $W^{(2)}$. 

We should mention that the 
expressions~(\ref{eq:gammahk}), (\ref{eq:gammacN}) for $\gamma$ are compatible with 
those obtained in \cite{romans, blumenhagen2a, Ahn:2012fz} for the finitely generated algebras
$\sWinf(1,2,\dots,N)$ with $N=2,3,4$,  and conjectured in~\cite{blumenhagen2a} for arbitrary 
$N$.

\subsection{Wedge Subalgebra}\label{sec:wedge}

As another  consistency check we can analyse whether the `wedge subalgebra' 
\cite{Bowcock:1991zk} 
of  $\sWinf[\mu]$ agrees indeed with $\shs[\mu]$, as must be the case if $\sWinf[\mu]$
is the Drinfel'd-Sokolov reduction of $\shs[\mu]$. Recall that the wedge subalgebra 
of $\sWinf[\mu]$ is defined
by restricting the modes $W^{s\, \alpha}_m$ to the wedge $|m|<s$, and taking the
limit $c\to\infty$. As can be seen from eq.~\eqref{central}, the $c\to\infty$ limit of 
$\gamma$ is zero unless we take 
$n_2$ to be proportional to the central charge; we can choose
\begin{equation}
n_2 = -\frac{c}{6}(\mu+1)(\mu-2) \ , 
\end{equation}
so that the structure constant $c_{22,2}$ equals
\begin{equation}\label{eq:wedgec222}
c_{22,2}= \frac{2}{\sqrt{3}}(1-2\mu) +\mathcal{O}(c^{-1}) \ .
\end{equation}
Note that this then reproduces the result of~\cite{Hanaki:2012yf}. With this normalisation 
convention we have checked that the other structure constants, that are determined by 
the OPEs (\ref{eq:ope22}) and (\ref{eq:ope23}) (with the coefficients given by 
(\ref{eq:soljac1}) -- (\ref{eq:soljac3})), agree indeed with those of $\shs[\mu]$; the details
are described in appendix~\ref{sec:shs}.


\section{Dualities}


As in the bosonic case \cite{GG3}, the actual $\sW_{\infty}$ algebra only depends on 
$\gamma$ and $c$. However, since the map $\mu\mapsto \gamma(\mu,c)$ is not injective,
there are in general different values of $\mu$ that lead to the same $\gamma$, and hence
to the same algebra. Indeed, if we fix $\gamma$ and $c$, 
then (\ref{central}) leads to a quartic equation for $\mu$. This means that we have 
a $4$-fold equivalence of algebras
\be \label{eq:4w}
\sW_{\infty}[\mu_1]  \cong \sW_{\infty}[\mu_2] \cong \sW_{\infty}[\mu_3] \cong \sW_{\infty}[\mu_4] \ ,
\ee
where the relation between the four parameters takes the remarkably simple form
\begin{align}\label{eq:4mu}
\mu_1 &= \mu\ ,& \mu_2 &=  \frac{c-c\mu}{c+3\mu}\ , & \mu_3 &= \frac{c+3\mu}{3\, (\mu-1)}\ ,
& \mu_4 & =  -\frac{c}{3\mu}\ .
\end{align}
Note that these relations break the classical $\mu \mapsto 1-\mu$ symmetry (that is 
obvious from the definition of ${\rm shs}[\mu]$, see (\ref{shsdef})) at finite $c$; this
is analogous to what happened in the bosonic analysis of \cite{GG3}, where the
$\mu\mapsto -\mu$ symmetry was similarly broken.

It is also useful to think about these identifications in the 
$(N,k)$-parametrisation, i.e.\ in terms of the Kazama-Suzuki  cosets 
\be
\frac{\SU(N+1)_k\times \SO(2N)_1}{\SU(N)_{k+1}\times \U(1)_\kappa} \ .
\ee
Then we have
\begin{align}
N_1 &= N\ ,&  k_1 &= k\ , \label{v1} \\
N_2 &= k \ ,& k_2 &= N\ , \label{v2} \\
N_3 &= -\frac{N}{N+k+1}\ ,&  k_3 &= -\frac{k}{N+k+1}\ ,\label{v3} \\
N_4 &= -\frac{k}{N+k+1}\ ,& k_4 &= -\frac{N}{N+k+1}\ . \label{v4}
\end{align}
Note that the equivalence between (\ref{v1}) and (\ref{v2}) (and similarly between
(\ref{v3}) and (\ref{v4})) is the familiar
level-rank duality of the Kazama-Suzuki models \cite{Gepner:1988wi}. On the other hand,
the relation between (\ref{v1}) and (\ref{v3}) explains the agreement of symmetries
between the 't~Hooft limit of the Kazama-Suzuki models, and the higher spin theory 
based on ${\rm shs}[\lambda]$ with
\be\label{lambdadef}
\lambda = \frac{N}{N+k+1} \ .
\ee
Indeed, (\ref{v1}) is the standard Kazama-Suzuki coset which is equivalent to 
(\ref{v3}), and hence to 
\be\label{claim}
\frac{\SU(N+1)_k\times \SO(2N)_1}{\SU(N)_{k+1}\times \U(1)_\kappa} \cong 
\sW_{\infty} [\lambda] \qquad \hbox{for $c=c_{N,k}$} \ ,
\ee
where we have used the dictionary $\mu_3=-N_3$. In the 't~Hooft limit, this 
proves that the two dual theories have equivalent symmetries, but (\ref{claim})
is actually a stronger statement since it applies also to finite $N$ and $k$. 

\subsection{Analytic Continuation}

Given the detailed understanding of the structure of the algebra, we can also
ask how the conformal dimensions of the various representations behave
in the `semiclassical' regime, i.e.\ for  large $c$. This 
analysis is of significance in order to determine which of the states of the CFT should
correspond to perturbative or non-perturbative higher spin excitations, respectively, see
\cite{GG3}. 


First we note that a generic $\sW_{\infty}[\mu]$ algebra has  four minimal
representations, since (\ref{eq:hresult}) has always four solutions for $h$. Plugging
in the expression for $\gamma$ in terms of $\mu$ and $c$, the four solutions are 
\be\label{4h}
h_1 = \frac{\mu}{2} \ , \qquad
h_2 = \frac{c (1-\mu)}{2 (c+3\mu)} \ , \qquad
h_3 = - \frac{c}{6 \mu} \ , \qquad
h_4 = - \frac{ (c + 3 \mu)}{6 (1 -\mu)} \ . 
\ee
As an aside, we can also write these formulae in terms of $(N,k)$, where they take the form
\begin{equation}\label{eq:4hmod}
h_1 = -\frac{N}{2}\ ,\quad
h_2 = -\frac{k}{2}\ ,\quad
h_3 = \frac{k}{2(N+k+1)}\ ,\quad
h_4 = \frac{N}{2(N+k+1)}\ .
\end{equation}
If we fix $\mu$ and consider the semi-classical limit ($c\rightarrow \infty$), two
of the solutions in (\ref{4h}), namely $h_1$ and $h_2$, are `perturbative' (since they
remain finite in this limit), while two solutions, namely $h_3$ and $h_4$,
are `non-perturbative' --- they are proportional to $c$ and go to
$-\infty$ in this limit (for $0<\mu<1$).

According to \cite{GG3}, the semiclassical limit is obtained by 
working with the higher spin theory based on $\mathfrak{sl}(N+1|N)$, and taking
$c\rightarrow \infty$ while keeping $N$ fixed. In order to understand what
happens in this limit, we should write 
the conformal dimensions of the two minimal representations of the
coset CFT (\ref{hNk}) in terms of $N$ and $c$; this leads to 
\begin{align}
& h({\rm f};0,N) = h (\bar{\rm f};0,-N) =  \frac{N}{2(N+k+1)}  = \frac{3N-c}{6 (N+1)}\cr
& h(0;{\rm f},-(N+1)) = h(0;\bar{\rm f},(N+1) ) = \frac{k}{2(N+k+1)} =  \frac{c}{6N} \ ,
\end{align}
where we have used (\ref{cNk}) to express $k$ in terms of $c$ (and $N$). 
These two solutions agree with $h_3$ and $h_4$ from (\ref{4h}, \ref{eq:4hmod})  for $\mu=-N$,
and hence are {\em both} non-perturbative. While this may sound somewhat
surprising at first, it actually ties in nicely with the results of \cite{GG3}.\footnote{We
thank Rajesh Gopakumar for the following suggestion.} To see this, recall that the
${\cal N}=2$ multiplet based on either $({\rm f};0)$ or $(0;{\rm f})$ actually contains
a scalar field that is quantised using the alternate $(-)$ quantisation, see Fig.~3 of 
\cite{Candu:2012jq}.\footnote{Incidentally, the same phenomenon also occurs in
one dimension higher, see \cite{Leigh:2003gk}.} From a bosonic point of view, we expect 
the fields with this 
alternate boundary condition to become non-perturbative in the semiclassical limit 
\cite{GG3}, and thus both ${\cal N}=2$ multiplets must show this behaviour,
in agreement with the above. It is tempting to speculate that the non-perturbative
characteristic of these fields is related to the
fact that, in $4$ dimensions, the scalar field with the alternate boundary condition
actually breaks the higher spin symmetry at finite $N$ \cite{Giombi:2011ya}, see also
\cite{Hartman:2006dy}.

Our result also suggests that the generalisation of the analysis of \cite{Castro:2011iw} to 
the supersymmetric case  should lead to more classical solutions; it would be very 
interesting to confirm this.


\section{Conclusion}

In this paper we have analysed the structure of the $\sWinf$ algebra that underlies
the higher spin -- CFT duality of \cite{Creutzig:2011fe}. This algebra is generated, in addition 
to the  ${\cal N}=2$ superconformal algebra, by exactly one ${\cal N}=2$
primary field for every integer spin $s\geq 2$. In particular, we have
solved the Jacobi identities arising from the first few OPEs, and we have found 
that the $\sWinf$ algebra is characterised by the central charge, as well as 
one free parameter that can be taken to describe the self-coupling $\gamma$ of the
spin $s=2$ primary field. 

From the point of view of the bulk AdS$_3$  theory, $\sWinf$ should be the quantum 
Drinfel'd-Sokolov reduction of the infinite superalgebra $\shs[\mu]$ that appears
in the Chern-Simons description of the higher spin theory. For $\mu=-N$, this DS algebra
is equivalent  \cite{Ito:1990ac} to a specific Kazama-Suzuki model 
\cite{Kazama:1988qp,Kazama:1988uz}, whose representation theory is known from the 
coset description. By comparing representations we have determined, as in 
\cite{GG3}, the exact relation between $\mu$ and the coupling constant 
$\gamma$ characterising $\sWinf$, see \eqref{central}. This identification
agrees with various results that had been previously determined in the
literature \cite{romans,blumenhagen2a}, and it is compatible with the requirement
that the wedge subalgebra of $\sWinf[\mu]$ reduces to $\shs[\mu]$.

It follows from \eqref{central} that there are generically $4$ different values
of $\mu$, see \eqref{eq:4mu}, that correspond to the same $\gamma$, and hence
define the {\em same} $\sWinf$ algebra. The $4$-fold equivalence of 
the $\sWinf[\mu]$ algebras \eqref{eq:4w} explains, among other things, the 
level-rank duality \cite{Gepner:1988wi} among the Kazama-Suzuki cosets,
see \eqref{v1} vs.\ \eqref{v2}. More importantly, it also establishes the 
equivalence of the quantum asymptotic symmetry algebra $\sWinf[\mu]$ of the higher 
spin theory on AdS$_3$, to the Kazama-Suzuki models in the duality of 
\cite{Creutzig:2011fe}.  As in \cite{GG3} this equivalence is actually true at
finite $N$ and $k$, and hence makes definitive predictions about the quantum
corrections of the higher spin theory. It would be very interesting to reproduce
these quantum corrections directly from the higher spin theory.

Among other things, our improved understanding of the quantum symmetry algebra
$\sWinf[\mu]$ also allows us to study the semi-classical (large $c$ at fixed $\mu=N$) 
behaviour of the two complex scalar fields that appear in the duality of 
\cite{Creutzig:2011fe}. Quite surprisingly, the conformal dimension of 
both dual fields is proportional to the central charge, thus suggesting that 
neither should be thought of as a perturbative scalar. Instead one should expect
that they have an interpretation in terms of 
`non-perturbative' classical solutions of the 
type found in \cite{Castro:2011iw} for the bosonic case; it would be very interesting
to check this in detail. A possible explanation for this non-perturbative  behaviour 
of the `scalar fields' is that all ${\cal N}=2$ matter multiplets involve
a scalar that is quantised in the alternate $(-)$ manner; such scalars 
turned out to be non-perturbative
in the bosonic analysis of \cite{GG3}. It would be very interesting
to understand this issue better, in particular, if there is a relation to the fact that 
in the AdS$_4$/CFT$_3$ duality the scalar field with the $(-)$ boundary condition 
breaks the higher spin symmetry at finite $N$.

\section*{Acknowledgements}

The work of CC and MRG is supported in parts by the
Swiss National Science Foundation. MRG thanks the Amsterdam String Theory
Workshop 2012 for hospitality during the final stages of this work.
We thank Rajesh Gopakumar, Maximilian Kelm 
and Carl Vollenweider for useful discussions.


%


\appendix

\section{Commutation Relations}\label{sec:22com}




The $\mathcal{N}=2$ Virasoro algebra is generated by the energy momentum tensor $T(z)$, 
a $\U(1)$ current $J(z)$ and two fermionic currents $G^\pm(z)$. Their OPEs take the
familiar form
\begin{align}\label{eq:vir2}
T(z)T(w)&\sim \frac{c}{2(z-w)^4}+\frac{2T(w)}{(z-w)^2}+\frac{\partial T(w)}{z-w}\ , \notag \\
T(z) J(w)&\sim \frac{J(w)}{(z-w)^2}+\frac{\partial J(w)}{z-w}\ ,\qquad
T(z) G^\pm(w) \sim  \frac{3G^\pm(w)}{2(z-w)^2}+\frac{\partial G^\pm(w)}{z-w}\ , \notag \\
J(z)J(w)&\sim \frac{c}{3(z-w)^2}\ , \qquad  J(z) G^\pm(w)\sim \pm \frac{G^\pm(w)}{z-m} \ , \\
G^+(z)G^-(w)&\sim \frac{2c}{3(z-w)^3}+\frac{2J(w)}{(z-w)^2} +\frac{2T(w)+\partial J(w)}{z-w}\ , \quad G^\pm(z)G^\pm(w)\sim 0\ . \notag
\end{align}
%
The commutation relations of the corresponding modes are then 
\begin{align}
[L_m,L_n] &= (m-n)L_{m+n} +\tfrac{c}{12}m(m^2-1)\delta_{m+n,0}\ ,\notag  \\
[L_m,J_n] &= -n J_{m+n}\ , \qquad [L_m,G^\pm_r] = \left(\tfrac{m}{2}-r\right) G^\pm_{m+r}\ ,\notag \\
[J_m,J_n]&=\tfrac{c}{3}\delta_{m+n,0}\ ,\qquad [J_m,G^\pm_r] = \pm G^\pm_{m+r}\ ,\\
[G^+_r,G^-_s] &=2L_{r+s}+(r-s)J_{r+s} +\tfrac{c}{3}\left(r^2-\tfrac{1}{4}\right)\delta_{r+s,0} \ ,
\qquad  [G^\pm_r,G^\pm_s]=0\ .\notag
\end{align}
%
%
Similarly, the modes of the fields in \eqref{eq:primary2opes} satisfy
\begin{align}\label{WsNcom}
[L_m,W^{s\, 0}_n]&=[(s-1)m-n]W^{s\, 0}_{m+n}\ ,\quad 
[L_m,W^{s\, \pm}_r]=\left[\left(s-\tfrac{1}{2}\right)m-r\right] W^{s\, \pm}_{m+r}\ , \notag \\
[L_m,W^{s\, 1}_n]&=[s m-n]W^{s\, 1}_{m+n}\ ,\quad  [J_m,W^{s\, 0}_n]=0 \ ,\quad 
[J_m,W^{s\, \pm}_r]=\pm W^{s\, \pm}_{m+r}\ , \notag \\
[J_m,W^{s \, 1}_n] &= s m\, W^{s\, 0}_{m+n}\ , \quad
[G^\pm_r,W^{s\, 0}_n]= \mp W^{s\, \pm}_{r+n}\ ,\\
[G^\pm_r,W^{s\, \mp}_t]&= \pm [(2s-1)r-t] W^{s\, 0}_{r+t}+2 W^{s\, 1}_{r+t}\ , \quad
[G^\pm_r,W^{s\, \pm}_t]= 0\ , \notag \\
 [G^\pm_r, W^{s\, 1}_n] &= \tfrac{1}{2}  [ (2s+1)r-n] W^{s\, \pm}_{r+n}\ . \notag 
\end{align}
%




{
\allowdisplaybreaks
\noindent 
Finally, the complete commutators  $[W^{2\, \alpha}_{m},W^{2\, \beta}_{n}]$ 
corresponding to the OPEs~\eqref{eq:ope22} take the form
\begin{align*}
 [W^{2\, 0}_{m}\, ,\, W^{2\, 0}_{n}] =& 
 (m-n)\mathscr{A}^{[2]}_{m+n}+\tfrac{c}{12} m(m^2-1)\delta_{m+n,0} \\
 [W^{2\, 1}_{m}\, ,\, W^{2\, 1}_{n}] =& \tfrac{c}{48} m(m^2-1)(m^2-4)\delta_{m+n,0}
+(m-n)(2m^2-mn+2n^2-8)\mathscr{B}^{[2]}_{m+n}\\
{}&+(m-n)\left(\mathscr{B}^{[4]}_{m+n} -2 c_{22}^3 W^{3\, 1}_{m+n}\right) \\
 [W^{2\, 0}_{m}\, ,\, W^{2\, 1}_{n}] =& \mathscr{C}^{[4]}_{m+n}+
(2m-n)\left(\mathscr{C}^{[3]}_{m+n} -c_{22}^3 W^{3\, 0}_{m+n}\right)\\
{}&+(6m^2-3mn+n^2-4)\mathscr{C}^{[2]}_{m+n}+m(m^2-1)\mathscr{C}^{[1]}_{m+n} \\
 [W^{2\, +}_{r}\, ,\, W^{2\, -}_{s}] =& \left(\mathscr{D}^{[4]}_{r+s} -2c_{22}^3 W^{3\, 1}_{r+s}\right)
+(r-s)\left(\mathscr{D}^{[3]}_{r+s} - 3c_{22}^3 W^{3\, 0}_{r+s}\right)\\
{}&+\left(3r^2-4rs+3s^2-\tfrac{9}{2}\right)\mathscr{D}^{[2]}_{r+s}
+(r-s)\left(r^2+s^2-\tfrac{5}{2}\right)\mathscr{D}^{[1]}_{r+s}\\
{}&+ \tfrac{c}{12}\left( r^2-\tfrac{1}{4}\right)\left( r^2-\tfrac{9}{4}\right)\delta_{r+s,0} \\
[W^{2\, +}_{r}\, ,\, W^{2\, +}_{s}] =& \mathscr{E}^{[4]}_{r+s}\\
[W^{2\, 0}_{m}\, ,\, W^{2\, +}_{r}] =& \left(\Phi^{[7/2]}_{m+r}-c_{22}^3 W^{3\, +}_{m+r}\right)
+\left( \tfrac{3}{2}m - r\right) \Phi^{[5/2]}_{m+r}\\
{}&+ \left( 3m^2 - 2mr +r^2-\tfrac{9}{4}\right) \Phi^{[3/2]}_{m+r} \\
[W^{2\, +}_{r}\, ,\, W^{2\, 1}_{m}] =& \Psi^{[9/2]}_{r+m} + \left( 2r-\tfrac{3}{2}m \right)
\left(\Psi^{[7/2]}_{r+m} -c_{22}^3 W^{3\, +}_{r+m}\right)\\
{}&+ \left( 2r^2 - 2rm +m^2 -\tfrac{5}{2}\right) \Psi^{[5/2]}_{r+m}\\
{}&+ \left( 4r^3 -3r^2m +2rm^2 - m^3 - 9r +\tfrac{19}{4}m  \right)\Psi^{[3/2]}_{r+m}\ .
\end{align*}
Here $\mathscr{A}^{[s]}$, $\mathscr{B}^{[s]}$,  $\mathscr{C}^{[s]}$,
$\mathscr{D}^{[s]}$,  $\mathscr{E}^{[s]}$
$\Phi^{[s]}$, $\Psi^{[s]}$ are defined 
exactly as in the paper of Romans~\cite{romans}, except that his coupling constant 
$\kappa$ should be regarded as a free parameter and identified with $c_{22,2}$.


\section{Composite Fields}
\label{sec:composite}

In this appendix we give the explicit form of the composite $\mathcal{N}=2$ primaries
that appear in our ansatz for the OPEs~(\ref{eq:ope23}) and (\ref{eq:ope3}).
The main reason for doing so, besides fixing the normalisation of their structure constants,
is to provide some non-trivial expressions that can be checked  at
intermediate steps of the calculation.

In our conventions, the normal ordered product $O_1 O_2$ of two operators $O_1$ and $O_2$
is the coefficient of the $(z-w)^0$ term in the OPE $O_1(z)O_2(w)$,
see \cite{thelemansthesis} for some properties of this normal ordered product.
(In particular,  it is not associative.)
In the expressions below, a normal ordered product of multiple operators
is always nested from the right, i.e.\ 
$O_1 O_2 \cdots O_n O_{n-1} := (O_1(O_2(\cdots (O_{n-1}O_{n})\cdots)))$.
With these conventions, the explicit expressions for $A^4$, $A^5$, 
$P^2$ and $P^{\frac{5}{2}}$ are

\begin{align*}
A^{4}=&
\left(W^{2\, 0}\right)^2-
\tfrac{9 \left(22 c^2-3c+9\right)n_2}{c(c-1) (c+1) (c+6) (2 c-3) (5 c-9)}  J^4 +
\tfrac{18 (c-33)  n_2}{(c+1) (c+6) (2 c-3) (5 c-9)}JG^+G^- \\
{}&-
\tfrac{2
(c+3) \left(44 c^2-129 c+99\right)  n_2}{(c-1) (c+1) (c+6) (2 c-3) (5 c-9)}T^2 +
\tfrac{12 \left(22 c^2-3c+9\right) n_2}{(c-1) (c+1) (c+6) (2 c-3) (5 c-9)} TJ^2\\
{}&+
\tfrac{3 (c-33) c  n_2}{(c+1) (c+6) (2 c-3) (5 c-9)} G^+\partial G^- +
\tfrac{9 \left(4 c^c+3^2-51 c+54\right) n_2}{2 c (c+1) (c+6) (2 c-3) (5 c-9)} (\partial J)^2 \\
{}&-
\tfrac{18 (c-33)  n_2}{(c+1) (c+6) (2c-3) (5c-9)} \partial TJ -
\tfrac{3 (c-33) c n_2}{(c+1) (c+6) (2c-3) (5c-9)} \partial G^+G^- \\
{}&+
\tfrac{6 \left(108-108 c-6 c^2-25 c^3+3 c^4\right)  n_2}{c(c-1)(c+1) (c+6) (2c-3) (5c-9)} \partial^2JJ +
\tfrac{3 (45+11 c) c_{22,2}}{(5c-12) \left(c^2+18c-51\right)} J^2W^{2\, 0} \\
{}&+
\tfrac{18 (c-33) (c-1)c_{22,2}}{(c+3) (5c-12) \left(c^2+18c-51\right)} JW^{2\, 1} -
\tfrac{2 \left(594-477 c+96 c^2+11 c^3\right) c_{22,2}}{(c+3) (5c-12) \left(c^2+18c-51\right)} TW^{2\, 0}\\
{}&+
\tfrac{3 (c-33) (c-1) c  c_{22,2}}{2 (c+3) (5c-12) \left(c^2+18c-51\right)} G^-W^{2\, +}-
\tfrac{3 (c-33) (c-1) c  c_{22,2}}{2 (c+3) (5c-12) \left(c^2+18c-51\right)} G^+W^{2\, -}  \\
{}&
- \tfrac{18 c_{22,3}}{7c-18} JW^{3\, 0}
+ \tfrac{2 c  c_{22,3}}{7c-18} W^{3\, 1} -
\tfrac{3 \left(27-42 c+c^2+2 c^3\right) n_2 }{(c+1) (c+6) (2c-3) (5c-9)} \partial^2T  \\
{}& -
\tfrac{3 (c-1) \left(108-87 c+10 c^2+c^3\right) c_{22,2} }{4 (c+3) (5c-12) \left(c^2+18c-51\right)} \partial^2W^{2\, 0}
-
\tfrac{(c-33) (c-6) n_2 }{2 (c+1) (c+6) (2c-3) (5c-9)} \partial^3 J
\ ,
\end{align*}
\begin{align*}
A^{5}=&W^{2\, 0}W^{2\, 1}+
\tfrac{(5c-24) }{8 c} W^{2\, +}W^{2\, -}-
\tfrac{6 }{c} J\left(W^{2\, 0}\right)^2   +
\tfrac{3 (c-4) \left(-42+12 c+c^2\right) c_{22,3} }{4 c \left(c^2+26c-75\right)}\partial^2W^{3\, 0}\\
{}&-
\tfrac{27 (11c-64)  c_{22,3}}{4 c\left(c^2+26c-75\right)} J^2W^{3\, 0}+
\tfrac{3 \left(-72-47 c+5 c^2\right)  c_{22,3}}{4 c \left(c^2+26c-75\right)} JW^{3\, 1}+
\tfrac{9 \left(72-81 c+17 c^2\right) c_{22,3}}{4 c \left(c^2+26c-75\right)} TW^{3\, 0}\\
{}&-
\tfrac{9 \left(200-103 c+17 c^2\right) c_{22,3}}{16 c \left(c^2+26c-75\right)} G^-W^{3\, +}+
\tfrac{9 \left(200-103 c+17 c^2\right) c_{22,3}}{16 c \left(c^2+26c-75\right)} G^+W^{3\, -}\\
{}&+
\tfrac{(5c-24) c_{22,3} }{8 c} \partial W^{3\, 1}
+
\tfrac{9 \left(-576-1956 c+257 c^2\right)c_{22,2}}{2 c (5c-12) (7c+6) \left(2c^2+9c-36\right)}  
J^3 W^{2\, 0} \\
{}& +
\tfrac{3 \left(-15552+4968 c+15192 c^2-2397 c^3+64 c^4\right) c_{22,2}}{2 c (c+3) (5c-12) (7c+6) \left(2c^2+9c-36\right)} J^2W^{2\, 1} +
\tfrac{3 (c-15) (5c-24)  c_{22,2}}{8 c (c+3) (5c-12)} J\partial W^{2\, 1}
 \\
{}& -
\tfrac{3\left(108864-13608 c-11178 c^2+3195 c^3-1065 c^4+242 c^5\right)  c_{22,2}}{16 c (c+3) (5c-12) (7c+6) \left(2c^2+9c-36\right)} J\partial^2W^{2\, 0}\\
{}&+
\tfrac{\left(139968-69336 c-42174 c^2+37809 c^3-6951 c^4-174 c^5+8 c^6\right) c_{22,2}}{8c (c+3) (5c-12) (7c+6) \left(2c^2+9c-36\right)}  \partial^2W^{2\, 1}\\
{}& -
\tfrac{3 \left(-46656+193320c-46782 c^2-16179 c^3+2333 c^4+414 c^5\right)  c_{22,2}}{16 c (c+3) (5c-12) (7c+6) \left(2c^2+9c-36\right)} \partial J\partial W^{2\, 0}\\
{}&-
\tfrac{3 \left(-31104-25056 c+22878 c^2-10599 c^3+656 c^4\right)c_{22,2}}{8 c (c+3) (5c-12) (7c+6) \left(2c^2+9c-36\right)}  JG^-W^{2\, +} \\
 {}&-
\tfrac{3 \left(62208-19872 c-19080 c^2+13872 c^3-3061 c^4+58 c^5\right) c_{22,2}}{8 c (c+3) (5c-12) (7c+6) \left(2c^2+9c-36\right)} G^-\partial W^{2\, +}\\
{}&+
\tfrac{3 \left(-31104-25056 c+22878 c^2-10599 c^3+656 c^4\right)  c_{22,2}}{8 c (c+3)(5c-12) (7c+6) \left(2c^2+9c-36\right)} JG^+W^{2\, -}  \\
{}&+
\tfrac{3 (c-15) (5c-24) c_{22,2}}{8 c (c+3) (5c-12)} \partial JW^{2\, 1}
-
\tfrac{27 (c-1) (5c-24)  c_{22,2}}{8 c (c+3) (5c-12)} T\partial W^{2\, 0} 
\\
{}&-
\tfrac{3 \left(-31104+4752 c+11052 c^2-8349 c^3+899 c^4\right)  c_{22,2}}{2 c (c+3) (5c-12) (7c+6) \left(2c^2+9c-36\right)} TJ W^{2\, 0}\\
{}&-
\tfrac{\left(15552-26136 c+15570 c^2-2757 c^3+46 c^4\right) c_{22,2}}{4 c(5c-12) (7c+6) \left(2c^2+9c-36\right)} TW^{2\, 1} -
\tfrac{(c-15) (2c-3) (5c-24) c_{22,2} }{48 c (c+3) (5c-12)} \partial^3W^{2\, 0}\\
{}&-
\tfrac{3 \left(-77760+76248 c+41274 c^2-37695 c^3+7033 c^4\right)c_{22,2}}{8 c (c+3) (5c-12) (7c+6) \left(2c^2+9c-36\right)}  G^+G^-W^{2\, 0}\\
{}&+
\tfrac{3 \left(-46656+51624 c+25398 c^2-24915 c^3+3427 c^4+222 c^5\right) c_{22,2}}{8 c (c+3) (5c-12) (7c+6) \left(2c^2+9c-36\right)} G^+\partial W^{2\, -} \\
{}&-
\tfrac{3 \left(15552-31320 c+8442 c^2+6117 c^3-3656 c^4+315 c^5\right) c_{22,2}}{4 c (c+3) (5c-12) (7c+6)\left(2c^2+9c-36\right)} \partial T W^{2\, 0} \\
{}&+
\tfrac{\left(93312-42768 c-22572 c^2+15462 c^3-219 c^4+10 c^5\right)  c_{22,2}}{8 c(c+3) (5c-12) (7c+6) \left(2c^2+9c-36\right)} \partial G^-W^{2\, +}\\
{}&+
\tfrac{\left(139968+52488 c-3618 c^2-17667 c^3+879 c^4+850 c^5\right) c_{22,2}}{8 c (c+3) (5c-12) (7c+6) \left(2c^2+9c-36\right)} \partial G^+W^{2\, -}\\
{}& -
\tfrac{\left(264384-55080 c-79812 c^2+41211 c^3-4827 c^4+199 c^5\right)  c_{22,2}}{4c (c+3) (5c-12) (7c+6) \left(2c^2+9c-36\right)} \partial^2JW^{2\, 0}\\
{}&-
\tfrac{\left(5184-11880c+7182 c^2-975 c^3-67 c^4+10 c^5\right) n_2 }{12 (c-2) (c-1) c (c+6) (c+12) (2c-3)} \partial^3 T -
\tfrac{3 (c-12) \left(144-30 c+19c^2\right)  n_2}{8 (c-2) c (c+6) (c+12) (2c-3)}G^+\partial^2G^- \\
{}&-
\tfrac{\left(31104-34992 c+8856c^2+2268 c^3-618 c^4+25 c^5\right)  n_2}{2 (c-2) (c-1) c^2 (c+6) (c+12) (2c-3)} T\partial^2J\\
{}&+
\tfrac{27 \left(87 c^3-386 c^2+1512 c-576\right)  n_2}{4(c-2) (c-1) c^2 (c+6) (c+12) (2c-3)} J^2 G^+G^- -
\tfrac{3 \left(53c^2-630c-1152\right)  n_2}{4 (c-2) (c-1) (c+6) (c+12) (2c-3)} JG^+\partial G^- \\
{}&-
\tfrac{3\left(10368+7344 c-7020 c^2+1344 c^3+67 c^4\right) n_2}{4 (c-2) (c-1) c^2 (c+6) (c+12) (2c-3)} J\partial G^+G^- +
\tfrac{45 (11c-102)  n_2}{(c-2) (c-1) (c+6) (c+12) (2c-3)} TJ^3 \\
{}&-
\tfrac{9 \left(-1728+1656 c+102 c^2-439c^3+45 c^4\right)  n_2}{2 (c-2) (c-1) c^2 (c+6) (c+12) (2c-3)} T^2J-
\tfrac{3 \left(-864+1692 c-894 c^2+157 c^3\right)  n_2}{(c-2) (c-1) c (c+6) (c+12) (2c-3)} TG^+G^-\\
{}&+
\tfrac{9 \left(1728+4536 c-2982c^2+111 c^3+65 c^4\right)  n_2}{8 (c-2) (c-1) c^2 (c+6) (c+12) (2c-3)} \left(J'\right)^2J -
\tfrac{3 \left(-3456+5040c-1686 c^2-125 c^3+45 c^4\right)  n_2}{2 (c-2) (c-1) c (c+6) (c+12) (2c-3)} \partial T T \\
{}&+
\tfrac{9 \left(1728-1656 c-102 c^2-71 c^3+10c^4\right)  n_2}{2 (c-2) (c-1) c^2 (c+6) (c+12) (2c-3)} T'J^2 -
\tfrac{3 (c-3) \left(3456-1008 c-828 c^2+84 c^3+25 c^4\right)  n_2}{4 (c-2) (c-1) c^2 (c+6) (c+12) (2c-3)} \partial T\partial J \\
{}&-
\tfrac{\left(2592-756c-162 c^2+39 c^3+16 c^4\right) n_2}{(c-2) (c-1) c (c+6) (c+12) (2c-3)}  \partial G^+\partial G^- +
\tfrac{3 \left(6912+3780 c-828 c^2+55 c^3\right) n_2}{4 (c-2) (c-1) c (c+6) (c+12) (2c-3)} \partial^2JJ^2\\
{}&+
\tfrac{3 (c-15) (5c-24)  n_2}{4(c-1) c (c+6) (2c-3)} \partial^2J\partial J-
\tfrac{9 \left(-1728+2232 c-258 c^2-23 c^3+5 c^4\right)  n_2}{4 (c-2) c^2 (c+6) (c+12) (2c-3)} \partial^2TJ\\
{}& -
\tfrac{3 (c-3) \left(-1728+5112c-1542 c^2-135 c^3+22 c^4\right) n_2}{4 (c-2) (c-1) c^2 (c+6) (c+12) (2c-3)} \partial^2 G^+G^- +
\tfrac{9 (4c+3) (5c-24)  n_2}{2 (c-1) c^2 (c+6) (2c-3)} T\partial JJ\\
{}&+
\tfrac{\left(-67392+50760 c-2538c^2-465 c^3-26 c^4+5 c^5\right)  n_2}{8 (c-2) (c-1) c^2 (c+6) (c+12) (2c-3)} \partial^3JJ -
\tfrac{9 (4c+3) (5c-24)  n_2}{4 (c-1) c^2 (c+6) (2c-3)} \partial JG^+G^- \\
{}&-
\tfrac{\left(-41472+62208 c-32112 c^2+4050 c^3+525 c^4-117c^5+2 c^6\right) n_2 }{16 (c-2) (c-1) c^2 (c+6) (c+12) (2c-3)} \partial^4 J \\
{} &-
\tfrac{27 (11c-102)  n_2}{(c-2) (c-1) c (c+6)(c+12) (2c-3)} J^5
\ ,
\end{align*}
\begin{align*}
P^2=&
W^{2\, 0}P
-
\tfrac{3 (3+5 c-12 h) w_2}{(c-6 h) (c+3-6h) (c-1-4h)} J^2P
-
\tfrac{2 \left(9 c-5 c^2-18 h-6 c h+8 c^2 h-48 c h^2\right) w_2}{(c-6 h) (c-1-4h) (1+2 h) (c-3+12h)} TP
\\
{}& 
+
\tfrac{18 (c-1) \left(-c+3 h+5 c h+c^2 h-6 h^2\right)  w_2}{(c-6 h) (c+3-6h) (c-1-4h) 
h (1+2 h) (c-3+12h)} J\partial P\\
{}&
-
\tfrac{3 (c-1) (c-12 h) \left(-c+c h-6 h^2\right) w_2}{(c-6 h) (c+3-6h) (c-1-4h) h (1+2 h) (c-3+12h)} G^+ P^-\\
{}&+
\tfrac{3 \left(27-33 c+3 c^2+3 c^3+78 c h-14 c^2 h-468 h^2
+84 c h^2+32 c^2 h^2-384 c h^3+1152 h^4\right)w_2}{(c-6 h) (c+3-6h) 
(c-1-4h) (1+2 h) (c-3+12h)}\partial JP\\
{}&-
\tfrac{3 (c-1) \left(-9 c+4 c^2+c^3+12 c h-72 h^2\right) w_2 }{2 (c-6 h) (c+3-6h) 
(c-1-4h) h (1+2 h) (c-3+12h)} \partial^2 P
\ ,
\end{align*}
and
\begin{align*}
P^{\frac{5}{2}}=&
W^{2\, 0} P^-
-
h W^{2\, -} P
+
\tfrac{3 \left(-3+5 c+12 h-26 c h-36 h^2+36 c h^2+72 h^3\right) w_2}{(c-6 h) 
h (1+2 h) (c-3+6h) (c-3+12h)} J^2 P^-\\
{}&-
\tfrac{6 \left(-3+5 c+12 h-26 c h-36 h^2+36 c h^2+72 h^3\right) w_2}{(c-6 h) 
(1+2 h) (c-3+6h) (c-3+12h)}  J G^- P \\
{}& -
\tfrac{9 (c-1) \left(-c+3 c h+6 h^2\right) w_2}{(c-6 h) h (1+2 h) (c-3+6h) (c-3+12h)} J \partial P^-\\
{}&+
\tfrac{(3+c) (3h-1) w_2}{h (c-3+6h) (c-3+12h)} T P^-
+
\tfrac{3 (c-1) (2h-1) \left(-6+c+6 h+5 c h+18 h^2\right) w_2}{(c-6 h) h (1+2 h) 
(c-3+6h) (c-3+12h)} G^- \partial P\\
{}&-
\tfrac{3 \left(6+5 c-3 c^2-78 h+7 c h+7 c^2 h+90 h^2-36 c h^2+2 c^2 h^2
+252 h^3+36 c h^3-288 h^4\right)w_2}{2 (c-6 h) h (1+2 h) (c-3+6h) (c-3+12h)}
\partial J P^-\\
{}&+
\tfrac{2 \left(-18+24 c-8 c^2-3 c h+5 c^2 h+234 h^2-156 c h^2+6 c^2 h^2-108 h^3+36 c h^3-432 h^4\right) w_2}{(c-6 h) (1+2 h) (c-3+6h) (c-3+12h)}
\partial G^- P\\
{}&+
\tfrac{3 (c-1) (3+c) (-2+c+2 h) w_2 }{2 (c-6 h) h (1+2 h) (c-3+6h) (c-3+12h)} \partial^2 P^-
\ .
\end{align*}
}

\section{Structure Constants of {$\shs[\mu]$}}
\label{sec:shs}

The structure constants of $\shs[\mu]$ were computed in a very explicit form in
\cite{Fradkin:1990qk}; to explain the basis that was used there, recall that
the bosonic subalgebra of $\shs[\mu]$ equals
\begin{equation}\label{eq:shs_bos}
\shs[\mu]_{0} \simeq \hs[1-\mu]\oplus\hs[\mu]\oplus \mathbb{C}\ ,
\end{equation}
where $\hs[\mu]$ is defined as
\begin{equation}\label{eq:hsdef}
\hs[\mu]\oplus\mathbb{C}=\frac{U(\mathfrak{sl}(2))}{\langle C^{\mathfrak{sl}}
-\frac{\mu^2-1}{4}\rangle}\ .
\end{equation}
Using the same conventions as in \cite{Gaberdiel:2011wb}, we denote the standard basis for 
$\hs[1-\mu]$ by $T^j_m$, and the standard basis for $\hs[\mu]$ by $U^j_m$. Furthermore,
the generator for the $\mathfrak{u}(1)$ factor in~\eqref{eq:shs_bos} is denoted by $v$. 
The fermionic generators of $\shs[\mu]$ have eigenvalue $\pm 1$ under 
the adjoint action of $v$; we denote the generators with eigenvalue $+1$ by 
$\Psi^{j}_r$,  $|r|\leq j$, $j=\tfrac{1}{2}, \tfrac{3}{2},\dots$,  and 
those with eigenvalue $-1$ by $\bar{\Psi}^j_r$. 
The commutation relations of $\shs[\mu]$ can then be written as
\begin{align}\notag
 [T^j_m,T^{j'}_{m'}] & =\sum_{j'',m''} f^{jj'j''}_{TTT}C^{jj'j''}_{mm'm''} T^{j''}_{m''}\ ,&
[U^j_m, U^{j'}_{m'}] & =\sum_{j'',m''} f_{UUU}^{jj'j''}C^{jj'j''}_{mm'm''} U^{j''}_{m''}\ ,\\ \label{eq:hscom1}
[T^j_m,\Psi^{j'}_{r'}] & =\sum_{j'',r''} f_{T\Psi\Psi}^{jj'j''}C^{jj'j''}_{mr'r''} \Psi^{j''}_{r''}\ , &
[T^j_m, \bar{\Psi}^{j'}_{r'}] & =\sum_{j'',r''} f_{T\bar{\Psi}\bar{\Psi}}^{jj'j''}C^{jj'j''}_{mr'r''} 
\bar{\Psi}^{j''}_{r''}\ ,\\ \notag
[U^j_m,\Psi^{j'}_{r'}] & =\sum_{j'',r''} f_{U\Psi\Psi}^{jj'j''}C^{jj'j''}_{mr'r''} \Psi^{j''}_{r''}\ , &
[U^j_m, \bar{\Psi}^{j'}_{r'}] & =\sum_{j'',r''} f_{U\bar{\Psi}\bar{\Psi}}^{jj'j''}C^{jj'j''}_{mr'r''} 
\bar{\Psi}^{j''}_{r''}\ ,
\end{align}
together with
\begin{equation}\label{eq:hscom2}
\{\Psi^j_r, \bar{\Psi}^{j'}_{r'}\}=\sum_{j'',m''} C^{jj'j''}_{rr'm''} 
\left( f_{\Psi\bar{\Psi}T}^{jj'j''} T^{j''}_{m''}  + f_{\Psi\bar{\Psi}U}^{jj'j''} U^{j''}_{m''}\right)\ ,
\end{equation}
where $C^{jj'j''}_{mm'm''}$ are the $\mathfrak{sl}(2)$ Clebsch-Gordan coefficients and 
the structure constants $f^{jj'j''}_{AA'A''}$ are those given in  \cite{Fradkin:1990qk};
explicitly they are\footnote{Note that there is a typo (wrong sign) in eq.~(27d) of 
\cite{Fradkin:1990qk}.} 
\begin{align}
f^{jj'j''}_{TTT} &= (1-\epsilon^{jj'j''}) F^{jj'j''}_{000}(1-\mu)\ ,&
f^{jj'j''}_{UUU} &= (1-\epsilon^{jj'j''}) F^{jj'j''}_{000}(-\mu)\ ,\\
f^{jj'j''}_{T\Psi\Psi} &= -\epsilon^{jj'j''} f^{jj'j''}_{T\bar{\Psi}\bar{\Psi}}\ ,&
f^{jj'j''}_{T\bar{\Psi}\bar{\Psi}} &= +F^{jj'j''}_{0\, 1/2\,1/2}(-\mu)\ ,\\
f^{jj'j''}_{U\Psi\Psi} &= -\epsilon^{jj'j''} f^{jj'j''}_{U\bar{\Psi}\bar{\Psi}}\ ,&
f^{jj'j''}_{U\bar{\Psi}\bar{\Psi}} &= -\epsilon^{jj'j''}F^{jj'j''}_{0\,-1/2\,-1/2}(1-\mu)\ ,\\
f^{jj'j''}_{\Psi\bar{\Psi}T}&=-\epsilon^{jj'j''}F^{jj'j''}_{1/2\,-1/2\,0}(1-\mu)\ ,&
f^{jj'j''}_{\Psi\bar{\Psi}U}&=-F^{jj'j''}_{-1/2\,1/2\,0}(-\mu)\ ,
\end{align}
where $\epsilon^{jj'j''}=(-1)^{j+j'-j''}$.
The symbols $F^{jj'j''}_{kk'k''}(\nu)$ were defined in~\cite{Fradkin:1990qk} as a deformation of 
the $6j$-symbols; their arguments must satisfy $|j-j'|\leq j''\leq j+j'$, $k''=k+k'$, 
$|k|\leq j$, $|k'|\leq j'$, $|k''|\leq j''$.
For this range of parameters one has
\begin{align*}
&F^{jj'j''}_{kk'k''}(\nu)=\sqrt{2j''+1}\Delta^{jj'j''}\sum_t (-1)^t \prod_{p=1}^{j+j'-j''-t}(\nu-j''+k''-p)\prod_{q=1}^{t}(\nu+j''+k''+q)\\
&\times \frac{\sqrt{ (j+k)!(j-k)!(j'+k')!(j'-k')!(j''+k'')!(j''-k'')!}}{t!(j+j'-j''-t)!(t+j''-j-k')!(t+j''-j'+k)!(j-k-t)!(j'+k'-t)!}\ ,
\end{align*}
where 
\begin{equation*}
\Delta^{jj'j''} = \sqrt{  \frac{(j+j'-j'')!(j+j''-j')!(j'+j''-j)!}{(j+j'+j''+1)!  }}\ .
\end{equation*}

In order to compare this basis to the wedge algebra of $\sW_{\infty}[\mu]$ it is convenient to
rescale all generators by the factor 
\begin{equation}
\alpha^j_m = \sqrt{\frac{(j-m)!(j+m)!}{(2j)!}}\ ;
\end{equation}
the rescaled generators will be denoted by small letters, e.g.\ 
$T^j_m = \alpha^j_m t^j_m$, etc. Then we have the identifications
{
\allowdisplaybreaks
\begin{align*}
J_0&=v\ ,&
G^+_r &= \sqrt{2} \psi^{1/2}_r\ , \\
G^-_r &= \sqrt{2} \bar{\psi}^{1/2}_r\ ,&
T_m &= -\frac{t^1_m+u^1_m}{\sqrt{2}}\ ,\\
W^{2\, 0}_m &= \frac{(\mu+1)t^1_m+(\mu-2)u^1_m}{\sqrt{6}}\ ,&
W^{2\, +}_r &= -\sqrt{2}\psi^{3/2}_r\ ,\\ 
W^{2\, -}_r &= -\sqrt{2}\bar{\psi}^{3/2}_r\ ,&
W^{2\, 1}_m &= \frac{t^2_m+u^2_m}{\sqrt{2}}\ ,\\
W^{3\, 0}_m &= \frac{3\left[ (\mu+2)t^2_m+(\mu-3)u^2_m \right] }{5\sqrt{6}c_{22,3}}\ ,&
W^{3\, +}_r &= -\frac{3\psi^{5/2}_r}{\sqrt{5}c_{22,3}}\ ,\\
W^{3\, -}_r &= -\frac{3\bar{\psi}^{5/2}_r}{\sqrt{5}c_{22,3}}\ ,&
W^{3\, 1}_m &= \frac{3(t^3_m+u^3_m)}{2\sqrt{5}c_{22,3}}\ ,\\
W^{4\, 0}_m &= \frac{3\sqrt{3}\left[ (\mu+3)t^3_m +(\mu-4)u^3_m\right]}{14\sqrt{5}c_{22,3}c_{23,4}}\ ,&
W^{4\, +}_r &= - \frac{3\sqrt{6}\psi^{7/2}_r}{\sqrt{35}c_{22,3}c_{23,4}}\ ,\\
W^{4\, -}_r &= - \frac{3\sqrt{6}\bar{\psi}^{7/2}_r}{\sqrt{35}c_{22,3}c_{23,4}}\ ,&
W^{4\, 1}_m &= \frac{3\sqrt{6}(t^4_m+u^4_m)}{2\sqrt{35}c_{22,3}c_{23,4}}\ .
\end{align*}
}
We have checked that, with these identifications and to the extent to which we have determined 
the commutation relations of $\sW_\infty[\mu]$, the wedge subalgebra of $\sW_\infty[\mu]$
agrees indeed with $\shs[\mu]$. Note that since $c_{22,2}=\sqrt{\gamma}$ has 
(for $c\rightarrow \infty$) a branch point at $\mu=\tfrac{1}{2}$, we have to be careful about
the branch of the square root we choose; we have worked with (\ref{eq:wedgec222}) 
and restricted 
$\mu$ to $\mu<\tfrac{1}{2}$. Furthermore, we have absorbed $A^{(4)}$ 
into the definition of $W^{(4)}$, see eq.\ (\ref{eq:ope23}),  i.e.\ we have set $a_{23,4}=0$.



\end{document}